\documentclass{ws-procs975x65}

\newcommand{\gs}{{_>\atop^{\sim}}}

\newcommand{\ltsima}{$\; \buildrel < \over \sim \;$}
\newcommand{\simlt}{\lower.5ex\hbox{\ltsima}}                                  
\newcommand{\lum}{\rm erg~s$^{-1}$}
\newcommand{\cgs}{${\rm erg~cm}^{-2}\ {\rm s}^{-1}$} 
\newcommand\arcsec{\mbox{$^{\prime\prime}$}}% 
\def\heao1{{\it HEAO-1\/}}

\def\asca{{\it ASCA\/}}

\def\sax{{\it BeppoSAX\/}}
\def\chandra{{\it Chandra\/}}
\def\xmm{{XMM-{\it Newton\/}}}

\def\apj{{\it ApJ}}
\def\apjs{{\it ApJS}}
\def\aap{{\it A\&A}}
\def\aj{{\it AJ}}

\def\newastr{{\it New Astr.}}
\def\etal{{\it et al.}}
\begin{document}

\title{The HELLAS2XMM 1dF Survey: a window on exotic hard X-ray 
selected sources}
%\footnote{\uppercase{T}his work is supported by etc, etc.}}

\author{C. Vignali\footnote{\uppercase{W}ork partially supported by the 
\uppercase{I}talian \uppercase{S}pace \uppercase{A}gency under the contract 
\uppercase{ASI} \uppercase{I/R}/057/02.}}

\address{INAF--Osservatorio Astronomico di Bologna, \\
Via Ranzani, 1 -- I-40127 Bologna, Italy \\
%E-mail: l\_vignali@bo.astro.it}
E-mail: chris@astro.psu.edu}
\author{on behalf of the HELLAS2XMM collaboration\footnote{
\uppercase{A. B}aldi, 
\uppercase{M. B}rusa, 
\uppercase{N. C}arangelo, 
\uppercase{P. C}iliegi, 
\uppercase{F. C}occhia, 
\uppercase{A. C}omastri, 
\uppercase{F. F}iore, 
\uppercase{F. L}a \uppercase{F}ranca, 
\uppercase{R. M}aiolino, 
\uppercase{G. M}att, 
\uppercase{M. M}ignoli, 
\uppercase{S. M}olendi, 
\uppercase{G.C. P}erola, 
\uppercase{L. P}ozzetti, 
\uppercase{S. P}uccetti 
and 
\uppercase{P. S}evergnini.
}}

%%%%%%%%%%%%%%%%%%%%%%%%%%%%%%%%%%%%%%%%%%%%%%%%%%%%%%%%%%%%%%
% You may repeat \author \address as often as necessary      %
%%%%%%%%%%%%%%%%%%%%%%%%%%%%%%%%%%%%%%%%%%%%%%%%%%%%%%%%%%%%%%

\maketitle

\abstracts{
Recent \chandra\ and \xmm\ surveys have confirmed that the cosmic X-ray 
background is mostly due to accretion onto super-massive black holes, 
integrated over cosmic time. 
Here we review the results obtained from the photometric 
and spectroscopic follow-up observations of the 122 X-ray sources detected 
by the HELLAS2XMM 1dF Survey down to a 2--10~keV flux of 
$\approx10^{-14}$~\cgs. 
In particular, we focus on the multiwavelength properties of a few intriguing 
classes of X-ray sources: high X-ray-to-optical flux ratio sources, 
Type~2 quasars, and XBONGs.}

\section{Introduction}
Hard X-ray surveys represent an efficient probe to unveil the 
super-massive black hole (SMBH) accretion activity, which is recorded in the 
cosmic X-ray background (XRB). 
The advent of hard X-ray (i.e., 2--10~keV) imaging instruments, from \asca\ 
to \sax\ and, more recently, \chandra\ and \xmm, has provided a dramatic 
advance in the field of X-ray surveys. 
The combination of deep/ultra-deep X-ray surveys with 
\chandra\cite{ale03,gia02} and the shallower surveys with \xmm\cite{bal02} 
has allowed to resolve \hbox{$\approx$~80\%} of the XRB 
in the \hbox{2--10~keV} band and to unveil classes of cosmic sources 
which were previously unknown or marginally represented by a few 
ambiguous and sparse cases. 
Within this context, the HELLAS2XMM\cite{bal02} survey plays an important 
role. Using suitable \xmm\ archival observations, 
this project aims at covering 4~sq.~degree of sky. 
Most of the fields are also characterized by \chandra\ observations, 
thus allowing for a more accurate determination of the \hbox{X-ray} source 
positions,\cite{bru03} hence possibly more reliable optical identification. 
At present, we have obtained optical 
photometric and spectroscopic follow-up of 122 sources 
in five \xmm\ fields, covering $\approx$~0.9~sq. degree (the HELLAS2XMM 
``1dF'' sample), down to a flux of $F_{\rm 2-10\ keV}\approx10^{-14}$\cgs, 
founding reliable optical spectroscopic redshifts and classification for 97 
of these sources.\cite{fio03} 
About two-third of the identified sources are broad-line AGNs; the remaining 
one-third are optically obscured AGNs whose nuclear optical emission is 
strongly reduced by dust and gas in the nuclear region and/or in the host 
galaxy. 
This class includes narrow-line (Type~2) AGNs, emission-line galaxies, and 
early-type galaxies with X-ray luminosity 
$\gs10^{42}$\lum\ (X-ray bright optically normal galaxies, XBONGs), 
strongly suggesting the presence of an active nucleus. 

In the following I focus on the properties of two intriguing classes of 
X-ray sources: %such as 
{\sf (a)} sources with high X-ray-to-optical flux ratio, 
which constitute the best candidate population suspected to host Type~2 
quasars, and 
{\sf (b)} XBONGs.

%%%%%%%%%%%%%%%%%
% X/O>10 sources
% Type 2 quasars
%%%%%%%%%%%%%%%%%
\par\noindent
{\sf (a)} About 20\% of the sources detected in recent hard X-ray surveys 
have a large X-ray \hbox{(2--10~keV)} to optical ($R$-band) 
flux ratio (X/0$>10$), i.e., ten times or more higher than typically observed 
in optically selected AGNs.\cite{fio03,ale01} 
X-ray hardness ratio analyses suggest that these objects are highly obscured. 
If they were at $z>1$, they could belong to the population of 
high-luminosity ($L_{\rm 2-10\ keV}\approx$~a few~$\times10^{44}$\lum), 
highly obscured ($N_{\rm H}\approx10^{22-23}$~cm$^{-2}$) AGNs, 
the so-called Type~2 
quasars postulated in the 
simplest versions of XRB models based on AGN unification schemes.\cite{com95} 
Unlike the faint sources found in ultra-deep \chandra\ and \xmm\ surveys, many 
of the extreme X/O sources in the HELLAS2XMM sample have $R$-band magnitudes 
\hbox{$<$24--25} and are therefore accessible to optical spectroscopy 
with 8-m class telescopes. 
%%%%%%%%%%%%%%%%%%%%%%%%%%%%%%%%%%%%%%%%%%%%%%%%%%%%%%%%
% STATUS X/O: 
% HELLAS2XMM: 13/28 spectr. identified (1dF Survey)=46%
% CDF-N+LH+SSA13 [from Fiore et al. 2003]: 9/38 (24%) at 
% approx. the HELLAS2XMM flux limit [3 in the LH from 
% photometric redshift] --> 6/35=17%
% Using Mignoli's ESO proposal and considering only 
% CDF-N sources: 3/59=5%
%%%%%%%%%%%%%%%%%%%%%%%%%%%%%%%%%%%%%%%%%%%%%%%%%%%%%%%%
At present, 13 out of the 28 HELLAS2XMM 1dF sources with X/O$>10$ (i.e., 
$\approx$~46\%, to be compared with 3/59~$\approx$~5\% in the CDF-N) 
have been spectroscopically identified (see Fig.~1, left panel); 
the properties of 8 of these sources place them into the long-sought 
class of Type~2 quasars\cite{fio03} at \hbox{$z\approx$~0.7--1.8}. 
%
%%%%%%%%%%%%%%%%%%%%%%%%%%%%%%%%%%%%%%%%%%%%%%%%%%%%%%%%%%%%%%
% Perola's fraction of Type 2 QSOs [Nh>10^22, L2-10>10^44]: 
% 16/57 with Zspec --> 28%
% 29/71 with Zphot --> 40%
%%%%%%%%%%%%%%%%%%%%%%%%%%%%%%%%%%%%%%%%%%%%%%%%%%%%%%%%%%%%%%
The presence of Type~2 quasars has been recently confirmed 
by direct X-ray spectral analysis.\cite{per04} 
The fraction of $L_{\rm 2-10\ keV}>10^{44}$\lum\ objects with 
$N_{\rm H}>10^{22}$~cm$^{-2}$ has been estimated to be 
$\approx$~28--40\%, and their surface density $\approx$~48/sq. degree 
at the flux limit of the HELLAS2XMM survey.\cite{per04}
%
%%%%%%%%%%%%%%%%%%%%%%%%%%%%%%%%%%%%%%%%%%%%%%%%%%%%%%%%%%%%%%
% X/O>10 sources without feasible optical spectroscopy (R>25): 
% Mignoli's sources and deep K-band photometry: Main results 
% from paper V
%%%%%%%%%%%%%%%%%%%%%%%%%%%%%%%%%%%%%%%%%%%%%%%%%%%%%%%%%%%%%%%
The redshift determination for the optically faintest ($R>25$) 
X/O$>10$ sources is not feasible via optical spectroscopy even with 
10-m class telescopes; using deep multi-band imaging to 
derive their redshifts would be time-consuming. 
However, deep near-infrared photometry ($K^{\prime}<21.5$) has been proven 
to be an efficient tool to obtain valuable information on the nature 
of this faint source population. 
Ten of the 11 HELLAS2XMM sources with X/O$>$10 observed in 
the $K^{\prime}$ band were detected 
($\langle K^{\prime}\rangle\approx$~18.4).\cite{mig04} 
Their optical-to-near-infrared colors are significantly redder than 
those of the field galaxy population, all of them being Extremely Red Objects 
(EROs; $R-K>5$) associated, in 7 out 10 cases, with bulge-dominated galaxies. 
Redshift estimates\cite{mig04} place these sources at $0.8<z<2.4$. 

%%%%%%%%%%%%%%%%%%%%%%%%%%%%%%%%%%%%%%%%%%%%
% Next shallow X-ray Surveys with XMM-Newton
%%%%%%%%%%%%%%%%%%%%%%%%%%%%%%%%%%%%%%%%%%%%
Moderately shallow, large-area surveys with \xmm\ (e.g., the extension of the 
HELLAS2XMM 1dF Survey to 4~sq.~degree and the 2~sq.~degree COSMOS), 
will provide further clues on the nature of the high X/O sources. 
Their redshift distribution, still highly incomplete, 
will also benefit from deep multi-filter imaging. 
The knowledge of their redshifts has strong consequences in the study of the 
energy density produced by accretion onto SMBH, since these objects are 
likely to carry the largest fraction of accretion power from the 
\hbox{$z\approx$~1--2} Universe. 

%%%%%%%%%
% XBONGs
%%%%%%%%%
\par\noindent
{\sf (b)} XBONGs, the first example detected in hard X-ray surveys being 
the object P3,\cite{fio00,com02}
are found at moderately low redshift ($z<1$) in both 
shallow and pencil-beam surveys.\cite{horn01,bar02} 
These objects do not show any evidence for 
AGN signatures in moderate-quality optical spectra, but their \hbox{X-ray} 
properties suggest they harbor powerful 
($L_{\rm 2-10\ keV}\approx10^{42-43}$\lum) AGNs. %obscured AGNs. 
Although the lack of clear AGN spectral features at optical 
wavelengths can be caused by a combination of wide slits, low-resolution, 
and low signal-to-noise ratio spectra having the effect of 
diluting the nuclear emission by the host galaxy 
starlight,\cite{moran02,seve03} 
a BL Lac nucleus can alternatively provide an 
apparent lack of emission lines, as recently observed.\cite{bru03} 
Moderately deep ($\approx$~21) \hbox{$K^{\prime}$-band} imaging of XBONGs 
is a powerful tool to provide constraints on the contribution of the AGN at 
longer wavelengths\cite{com04} and reveal the presence of 
complex morphologies (e.g., the double nucleus in Fig.~1, right panel).
%
%%%%%%%%%%%%%%%%%%%%%%%%%%%%%%%%%%%%%%%%%%%%%%%%%%%%%%%%%%%%%%%%%%%%%%
% Figure 1 - two panels: (a) optical spectrum of a Type 2 QSO;
%                        (b) K-band image of a morphologically 
%                          ``peculiar'' XBONG (double-nucleus);
\begin{figure}[t]
\parbox{0.48\textwidth}
{\psfig{figure=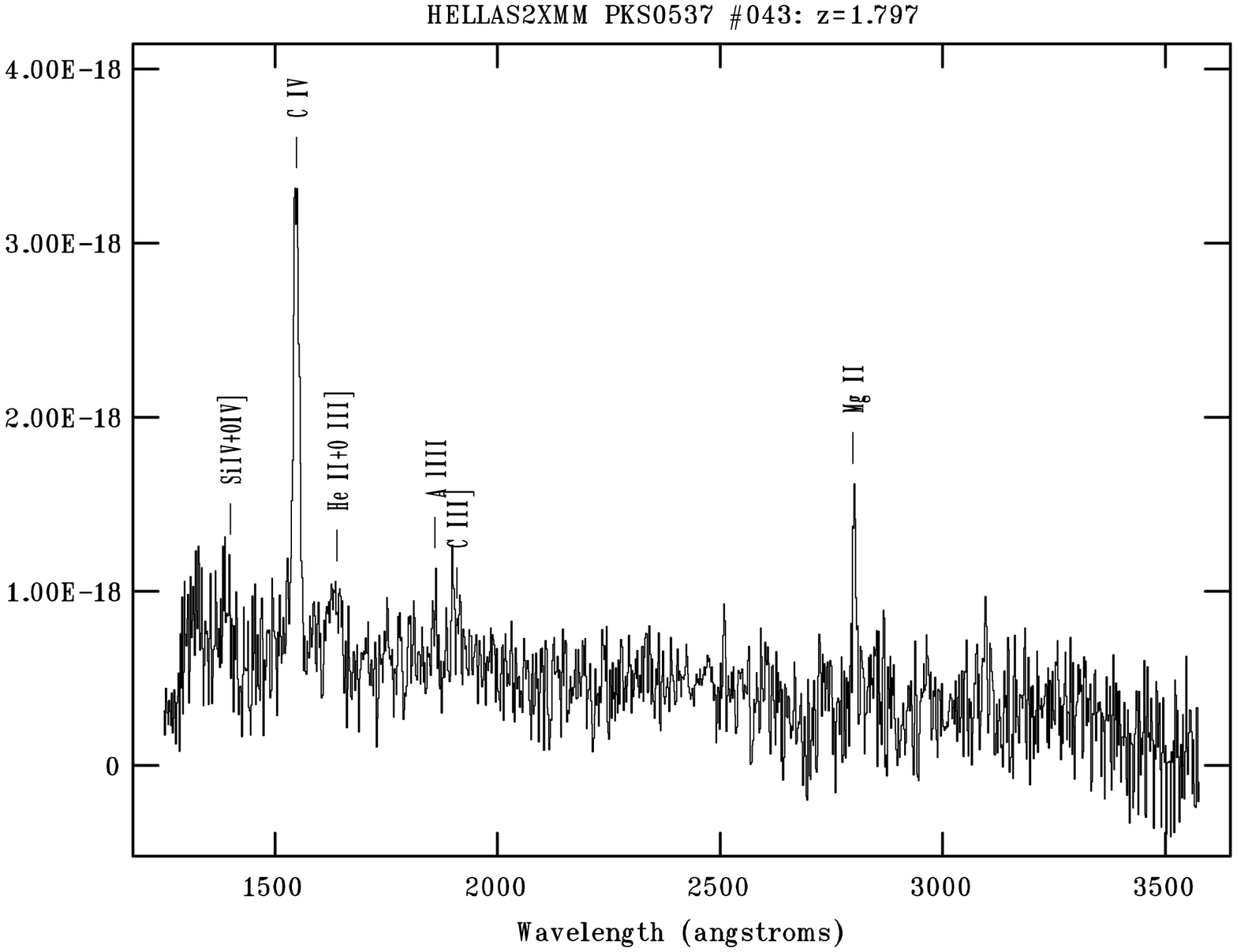,width=15pc,angle=90}}
\hspace{1.6cm}
\parbox{0.48\textwidth}
{\vskip -0.7cm \psfig{figure=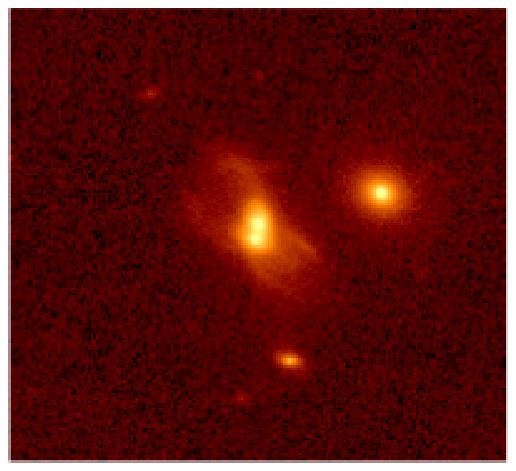,width=10pc,angle=0}}
\caption{{\it Left:} optical spectrum of a Type~2 quasar at $z=1.797$. 
{\it Right:} $K^{\prime}$-band image of a ``peculiar'' double-nucleus XBONG at $z=0.251$ 
(seeing $\approx$~0.5\arcsec\ FWHM).}
\label{fig1}
\end{figure}
%%%%%%%%%%%%%%%%%%%%%%%%%%%%%%%%%%%%%%%%%%%%%%%%%%%%%%%%%%%%%%%%%%%%%%

\end{document}